\documentstyle[pra,epsfig,floats,aps]{revtex}

\begin{document}

\title{Optimisation problems and replica symmetry breaking \\
in finite connectivity spin-glasses }
\author{R\'emi Monasson \cite{rm} }
\address{CNRS - Laboratoire de Physique Th\'eorique de l'ENS,
24 rue Lhomond, 75231 Paris cedex 05, France}
\date{\today}
\maketitle

\begin{abstract}
A formalism capable of handling the first step of hierarchical replica
symmetry breaking in finite-connectivity models is introduced.
The emerging order parameter is claimed to be a probability
distribution over the space of field distributions (or, equivalently
magnetisation distributions) inside the cluster of states. 
The approach is shown to coincide with the previous works
in the replica symmetric case and in the two limit cases $m=0$,
$m=1$ where $m$ is Parisi's break-point. As
an application to the study of optimization problems, the ground-state
properties of the random 3-Satisfiability problem are investigated and
we present a first RSB solution improving replica symmetric results.
\end{abstract}


\section{Introduction}

It is now commonly thought that spin-glasses may exhibit highly
interesting and non trivial features already at the mean-field level
\cite{sk,mpv}. This statement stems from the extensive studies
performed during the last twenty years on models with infinite
connectivity (IC), especially on the celebrated
Sherrington-Kirkpatrick (SK) model \cite{sk}.  Such models strongly
differ from realistic finite-dimensional systems on two points, that
is their large connectivity and the absence of any geometrical
underlying structure. As was realized some years ago, only the latter
aspect is intrinsic to mean-field theory while the irrealistic nature
of the connectivity may be cured 
\cite{viana,kanter,mezard,dom1,dom2,wong,gold}.
Mean-field spin-glasses with finite connectivity (FC) are of
importance for at least two reasons. First, they are expected to share
common properties with finite-dimensional physical systems that IC
models cannot exhibit \cite{prop}. Secondly, it is now well-known that
optimisation or decision problems can rigorously be mapped onto the
ground-state of spin-glass models with FC \cite{heidel}. The
understanding of complex optimisation problems \cite{garey}, which
would be of practical use for algorithmic design, therefore requires
the introduction of sophisticated techniques that were invented in the
IC models context, {\em e.g.}  replica symmetry breaking (RSB)
\cite{mpv}.  As regard to their potential interest, the investigation
of the spin-glass phase of FC models has attracted few attention in
the past years\cite{dom1,dom2,wong,gold}. This situation is probably
mainly due to the technical difficulties arising in the analytical
calculations.

In this paper, we show that the complexity of FC models with respect
to IC ones manifests itself through the emergence of new and richer
order parameters (Section II). We expose how to compute the
free-energy of FC models within the one-step RSB scheme in Section
III.  Section IV is devoted to the derivation of the saddle-point
equation for the functional order parameter.  Throughout
the paper, we check that the present formalism gives back the known
results for the RS theory of FC models and the RSB theory of IC
models.  Our approach can be applied to any Ising FC spin-glass and is
therefore of particular relevance to optimization problems.
As an illustration, we concentrate in Section V upon the so-called random 
3-Satisfiability problem (3-SAT),
of central importance in complexity theory \cite{garey,hayes}. Recent
theoretical interpretations \cite{court,long} of the numerical data
accumulated so far on 3-SAT \cite{nume} speak indeed for the existence
of a spin--glass phase, making 3-SAT a valuable testing ground for RSB
calculations. A first RSB solution is found and shown to
improve the RS result.

\section{Order Parameter for finite-connectivity models}

\subsection{Occupation densities and the multi-level gas picture}

Let ${\cal H}$ be a Hamiltonian depending on $N$ Ising spins $S_i$,
$i=1,\ldots ,N$ and on some quenched degrees of freedom.  To compute
the equilibrium properties, we resort to the replica method
\cite{mpv}. Once the system has been replicated $n$ times and the
disorder averaged out, we obtain an effective model of $2^n$-states
spins $\vec S _i = (S^1_i,S^2_i,\ldots ,S^n_i)$. In the absence of any
underlying geometry ({\em e.g.} lattice), the effective Hamitonian
${\cal H}_{eff}$ is invariant under any relabelling of the sites. As a
consequence, ${\cal H}_{eff}$ depends upon the spins {\em only}
through the $2^n$ occupation densities $c(\vec \sigma )$ defined as
the normalized fractions of sites $i$ such that $\vec S _i = \vec
\sigma$ \cite{long,eilat}. This is the very meaning of a mean-field
theory~: the effective Hamiltonian depends on the spins only through a
finite set of global fields rather than an extensive set of local
ones. For instance, the effective Hamiltonian of the SK 
model at inverse temperature $\beta$ \cite{sk,eilat} reads
\begin{eqnarray}
N \ {\cal H}_{eff}^{SK} &=& -\frac{\beta}{2 N}\sum_{ i < j}
\left( \sum _{a=1} ^n S_i ^a S_j ^a \right) ^2 \nonumber \\
&\simeq & -\frac{\beta N}{4}\sum_{
\vec \sigma _1 , \vec \sigma _2} c(\vec \sigma _1 )\; c(\vec \sigma _2 )
\; (\vec \sigma _1 . \vec \sigma _2)^2 
\qquad ,
\label{kur2sk}
\end{eqnarray}
up to $O(1)$ irrelevant terms.

In this picture, we may interpret any mean-field IC or FC replicated
system as a `gas' of $N$ particles living on $2^n$ interacting
levels. Each level is labelled by a vector $\vec \sigma$ comprised of
$n$ binary components $\sigma ^a =\pm 1$ and is filled in with $N.c(\vec
\sigma )$ particles. Taking into account the entropic contribution
coming from the combinatorial choices of the sites, we obtain the
general expression for the $n^{th}$ moment of the partition function
$Z$,
\begin{equation}
\overline{Z^n} = \int _0 ^1 \prod _{\vec \sigma }
dc(\vec \sigma ) \; \delta \left( \sum _{\vec \sigma }
c(\vec \sigma ) - 1 \right) \; e^{ -N \beta {\cal F}(\{ c\} ) }
\label{e1}
\end{equation}
where the bar denotes
the average over the disorder and the free-energy functional reads, to
the largest order in $N$,
\begin{equation}
{\cal F}(\{ c\} ) = {\cal H} _{eff} (\{ c\} ) + \frac{1}{\beta} \sum
_{\vec \sigma} c(\vec \sigma ) \ln c(\vec \sigma ) \qquad .
\label{e2}
\end{equation}
In the thermodynamical limit, the occupation densities $c(\vec\sigma
)$ are determined through the $2^n$ saddle-point equations
corresponding to the optimisation of the free energy functional 
${\cal F}$, 
\begin{equation}
c(\vec \sigma ) = \lambda (n) \ \exp \left( -\beta
\; \frac{\partial {\cal H} _{eff}}{\partial c (\vec \sigma )}
\right)
\label{kur1}
\end{equation}
where $\lambda (n)$ is 
to be chosen to ensure the normalization of the
$c$'s. All equilibrium properties can then be computed.

Let us briefly comment the pros and the cons of the above formalism
with respect to the usual formulation which involves overlaps between
spins belonging to different replicas.  Expressions (\ref{e1},\ref{e2})
always require to compute $2^n$ order parameters $c$. In the case of
IC models, most of them contain redundant information since only
$\frac 12 n(n-1)$ distinct overlaps are required. As we shall see in
the following, the structure of the order parameters of IC models is
indeed very simple and does not account for the whole spectrum of
mean-field disordered models. For FC systems, the present formalism
proves to be much more tractable than the usual formulation from the
analytical standpoint. It amounts to encode all overlaps in a concise
way through the generating function $c(\vec \sigma )$
\cite{rem1}. Another advantage of the present approach appears when
focusing on Hamiltonians with $p$-spins interactions ($p\ge 3$)
\cite{pspin}. The use of the occupation densities $c$'s
avoids the physically unclear introduction of Lagrange
parameters, that are necessary to define overlaps even in the IC case. 

\subsection{Replica symmetric Ansatz}

With a view to undertake RSB calculations, a short discussion of the
replica symmetric (RS) Ansatz is illuminating. The RS theory of FC
models was worked out ten years ago \cite{kanter,mezard} and may be
reformulated within the Thouless-Anderson-Palmer (TAP) framework
\cite{tap}. Inside the single RS state, the spins $S_i$ fluctuate
around their Gibbs averages $\langle S_i \rangle $. All relevant
information rest in the histogram $p_{rs} (h)$ of the effective fields
$h_i = \frac 1\beta \tanh ^{-1}( \langle S_i \rangle )$ \cite{kanter}.
The order parameter proves thus to be a function $p_{rs}$, belonging
to the space $V$ of the probability distributions over real numbers.
How are these results recovered in the present formalism ? RS
corresponds to the invariance of the saddle-point $c(\vec \sigma)$
under any permutation of its components. In other words, $c(\vec
\sigma)$ is a function of the magnetisation $s=\displaystyle{\sum
_{a=1}^n \sigma^a}$ only.  Inserting the above Ansatz into the
extremisation conditions of ${\cal F}(\{ c\} )$, one easily finds back
all thermodynamical quantities through the one-to-one correspondence
\begin{equation}
c(\vec \sigma ) = \int _R dh\; p_{rs}(h) \prod
_{a=1}^n \left( \frac{ e^{\beta h \sigma ^ a}}{e^{-\beta h} +
e^{\beta h}} \right) \qquad ,
\label{pwer1}
\end{equation}
or, equivalently,
\begin{equation}
c(s) = \int _R dh\; p_{rs}(h) 
\frac{ e^{\beta h s }}{(2 \cosh \beta h )^n} 
\qquad ,
\label{pwer2}
\end{equation}
where the symbol $\int _R $ stands for an integral over the whole
real axis. Note that the integral over $h$ in (\ref{pwer2}) converges
if the real part of the magnetisation $s$ is smaller than $n$ (in
absolute value). Therefore, when the number of replicas $n$ tends to
zero, the order parameter $c(s)$ may be analytically continued on the
imaginary axis. This is precisely what one needs to go back to
the field distribution 
\begin{equation}
p_{rs} (h) = \int _R \frac{ds}{2 \pi} \ c ( i s /\beta )
\ e^{ - i s h}
\qquad \qquad (n \to 0 ) \qquad .
\end{equation}
Let us underline that a drastic simplification takes place in IC
models. The field distribution $p_{rs}$ (and $c$) becomes
Gaussian; it is fully described by a
variance, the RS overlap $q$ \cite{mpv}, and the functional nature of
the order parameter is hidden.

\subsection{Replica symmetry broken Ansatz}

We now consider the first step of Parisi's hierarchical RSB scheme
\cite{mpv}, regardless of the possible existence of any new RSB
pattern in FC spin-glasses. Each replica $a$ is labelled by a couple
of integers $(b,\tilde b)$ according to the number $b$ of the block it
belongs to ($1\le b\le n/m$) and its position $\tilde b$ inside this
block ($1\le \tilde b \le m$). Following De Dominicis and Mottishaw
\cite{dom2}, we see that $c(\vec \sigma )$ is left unchanged under any
permutation inside the blocks and depends only on the block
magnetisations $\displaystyle{s_b = \sum _{\tilde b =1}^m \sigma
^{(b,\tilde b)}}$. To guess the structure of the order parameter, we
again resort to the TAP approach. Consider spin $S_i$. Due to the
presence of numerous states in the single cluster, the thermal average
$\langle S_i \rangle $ fluctuates from state to state.  Consequently,
the effective field $h _i$ is distributed according to a function
$\rho _i (h ) \in V$. We must keep in mind that $\rho _i$ is in turn a
random variable depending on the particular spin under consideration.  The
variety of $\rho$'s may be taken into account by introducing their
histogram, that is a functional ${\cal P}[ \rho ]$. This is a
normalized distribution over $V$~: $\int _V {\cal D} \rho \ {\cal P} [
\rho ] = 1$.  We thus look for ${\cal P}$ such that
\begin{equation} 
c ( \vec \sigma ) = 
\int _V {\cal D} \rho \ {\cal P} [\rho ] \; 
\prod_{b=1} ^{n/m} \int _R dh\; \rho (h)
\frac{e^{\beta h s_b}}{ (2 \cosh \beta h )^m}
\label{crsb}
\end{equation}
satisfies the saddle-point equations for ${\cal F}$.  The generic
order parameter ${\cal P}$ for FC models appears to be much more
complex than in the IC case. 

As a simple check of the above formalism, let us see how to find back
the RS theory. In the latter case, there exists a unique state. The
effective fields can therefore not fluctuate from `state' to `state'
and $\rho _i$ is simply a Dirac distribution in $h_i$. The fields $h_i$
fluctuate according to their distribution $p_{rs}$. Defining
\begin{equation}
{\cal P}_{rs} [ \rho ] = \int _R d\tilde h \; p_{rs} (\tilde h ) \
\delta \bigg[ \rho (h) - \delta ( h -\tilde h ) \bigg] \qquad ,
\label{rsfgt}
\end{equation}
we indeed find that the order parameter (\ref{crsb}) simplifies to the
RS expression (\ref{pwer1},\ref{pwer2}). In eq. (\ref{rsfgt}),
the symbol $\delta$ denotes the Dirac functional,  {\em i.e.}  the
product over all values of $h$ is omitted for simplicity.
It was first remarked by Wong and
Sherrington (in a different formalism) \cite{wong} that a
generalization of the above equation to the RSB case may be obtained
by replacing the inner Dirac distribution in eq.~(\ref{rsfgt}) with a
function to be optimised over.  This Ansatz may be correct if the
distributions $\rho $ with a non zero weight can be labelled by ({\em}
i.e. are not more `numerous' than) real numbers. In the generic case,
a full functional ${\cal P}$ is {\em a priori} needed.

How can expression (\ref{crsb}) be analytically continued to real
$n,m$? We first define $\nu (y)$ as the number of blocks $b$ of
magnetisations $s_b =y$, with $y=-m,-m+2 ,\ldots ,m-2
,m$. It is easy to check out on (\ref{crsb}) that $c(\vec \sigma )$
depends only on the set of $\nu (y)$'s, as expected from the
invariance of the order parameter under permutations between blocks
\cite{dom1}. The discrete nature of $y$ is merely due to the integer
value of $m$ and can be omitted to define an analytical continuation
of the order parameters. Consequently, $\nu (y)$ may be any function
on the range $-m \le y \le m$ satisfying the constraint
\begin{equation}
\int _{-m}^m dy \; \nu (y) =  \frac nm \to 0
\qquad ,
\label{hdw}
\end{equation}
 in the small $n$ limit.  Finally,
the order parameter $c$ becomes a functional over the set of all
possible functions $\nu$ and reads, from (\ref{crsb}),
\begin{equation} 
c [ \nu ] = 
\int _V {\cal D} \rho \ {\cal P} [\rho ] \;
\exp \left(  \int _{-m} ^m dy \; \nu (y)  \
\ln \left[ \int _R dh\; \rho (h) \frac{ e^{\beta h y}}
{(2 \cosh \beta h )^m} \right] \right)
\qquad .
\label{cnu}
\end{equation}
Notice that the RS case is recovered by injecting the order parameter
(\ref{rsfgt}) into (\ref{cnu}); one recovers (\ref{pwer2}) with $s=\int
_{-m} ^m dy \; \nu (y) \; y$.  In the generic RSB case, $c[\nu ]$
depends on the whole $\nu$ function and not only on its first moment
\cite{clown}. 

\section{Expression of the one-step free-energy functional}

\subsection{Methods to compute the free-energy}

Two procedures can be followed to access the thermodynamical
properties, depending on the starting point of the calculation. 

\begin{itemize}

\item{{\em From the saddle-point equation }:}
We inject the expression of the order parameter (\ref{cnu}) into
the saddle-point equation (\ref{kur1}). The resulting equation for
$P[\rho ]$ has to be solved. Then, one can compute the free-energy
(\ref{e2}), possibly using the saddle-point equation to simplify the
calculation (especially the entropic term whose calculation is uneasy).  

\item{{\em From the free-energy functional }:}
Since we know that the saddle-point equation is closed within the
one-step algebra, we may first compute the free-energy functional
${\cal F}$ given in (\ref{e2}) restricted to one-step order parameters
(\ref{crsb}). Once the analytical continuation of ${\cal F}$ to real
$m, n(\to 0)$ has been carried out, we obtain the saddle-point
equation for $P[\rho ]$ by differentiating ${\cal F}$ with respect to
the latter.

\end{itemize}

While both methods lead to the same result, the second one has an
important practical advantage. The (total) derivative of ${\cal F}$
with respect to any parameter is equal to the partial derivative,
while this is not necessarily true in the first procedure. This is of
little interest for most of the control parameters which appear in the
effective Hamiltonian only (it is usually easy to compute the partial
derivative of ${\cal H}_{eff}$ in (\ref{e2}) and to perform the
analytical continuation - see next paragraph). However, considerable
simplifications arise when computing the derivative of ${\cal F}$ with
respect to $m$. In the following, we shall therefore adopt the second
procedure.

\subsection{Energetic contribution}

According to the interpretation given in Section II, the effective
hamiltonian ${\cal H}_{eff}$ describes the interactions between the
different levels $\vec \sigma$. If the levels interact $K$ by $K$, the 
corresponding effective Hamiltonian typically reads
\begin{equation}
{\cal H}_{eff} \sim \sum _{\vec \sigma _1 ,\vec \sigma _2 , \ldots ,
\vec \sigma _K } c(\vec \sigma _1 ) c(\vec \sigma _2 ) \ldots
c(\vec \sigma _K ) \ I(\vec \sigma _1 ,\vec \sigma _2 , \ldots ,
\vec \sigma _K ) 
\qquad
\end{equation}
where the interaction function $I$ is invariant under global permutations
of both replicas and levels labels. For all usual IC or FC models, the 
computation of ${\cal H}_{eff}$ within the Ansatz (\ref{crsb}) as well as 
taking the limit $n\to 0$ do not present any difficulty.

As an illustration, we consider three examples~:
\begin{itemize}

\item{{\em the SK model} :} inserting $c(\vec \sigma )$ (\ref{crsb}) in
${\cal H}_{eff}$ (\ref{kur2sk}), the trace over $\vec \sigma$ can be 
straightforwardly carried out by writing 
\begin{equation}
(\vec \sigma _1 .\vec \sigma _2
)^2 = \frac 12 \ \frac{d^2}{dz^2} \bigg| _{z=0} \exp \left( z \sum _{a=1}^n 
\sigma _1 ^a \sigma _2 ^a \right) 
\qquad . 
\end{equation}
We obtain ~:
\begin{eqnarray}
\frac{1}{n} {\cal H}_{eff} ^{SK} &=& -\frac{\beta}{4n}
\frac{d^2}{dz^2} \bigg| _{z=0} \int _V {\cal D}\rho _1 {\cal D}\rho _2
\; {\cal P} [\rho _1 ] {\cal P} [\rho _2 ]  \; \left[
\int _R dh_1 dh_2 \right. \nonumber \\ 
& & \left. \rho _1 (h_1 ) \rho _2 (h_2 ) 
\left( \cosh z + \sinh z \tanh \beta h_1 \tanh \beta h_2 
\right) ^m \right] ^{n/m}
\nonumber \\
&=& -\frac{\beta}{4} \left( 1 - q_1 ^2 + m (q_1 ^2 - q_0 ^2) \right)
\qquad \qquad (n\to 0 ) \qquad ,
\label{enersk}
\end{eqnarray}
where
\begin{eqnarray}
q_0 &=& \int _V {\cal D} \rho {\cal P}[\rho ] \left( \int _R dh \rho
(h) \tanh \beta h \right) ^2 \nonumber \\ q_1 &=& \int _V {\cal D}
\rho {\cal P}[\rho ] \int _R dh \rho (h) (\tanh \beta h ) ^2 \qquad .
\label{poiu}
\end{eqnarray}
As expected for IC models, the functional
nature of the order parameter is indeed drastically washed out since
only the two moments $q_0 , q_1$ of ${\cal P}[\rho ]$ are relevant.

\item{{\em the Viana-Bray model} :} this model was introduced
as a FC version of the SK model \cite{viana}. The Viana-Bray
Hamiltonian reads
\begin{equation}
{\cal H}_{eff} ^{VB} = \frac {\alpha}{\beta} \ \left[ \ 1 - 
\sum _{\vec \sigma _1 , 
\vec \sigma _2 } c(\vec \sigma _1 ) \
c(\vec \sigma _2 ) \ \cosh \left( \beta \ 
\vec \sigma _1 . \vec \sigma _2 \right) \right]
\qquad ,
\label{vb}
\end{equation}
where $\alpha$ denotes the  mean connectivity per spin. We obtain
\begin{eqnarray}
\frac{1}{n} {\cal H}_{eff} ^{VB}&=& - \frac{\alpha}{\beta m} 
\int _V {\cal D} \rho _1 \; {\cal D}\rho _2 \; {\cal P}[\rho _1 ]
\; {\cal P}[\rho _2 ] \ln \left[ \int _R dh_1  dh_2
\right .\nonumber \\ && \left. \rho _1(h_1) \;\rho _2(h_2) 
\left( \cosh \beta + \sinh \beta \;\tanh \beta h_1 \tanh \beta h_2  \right)
^m \right] \qquad , 
\label{enervb}
\end{eqnarray}
which explicitly depends upon on the whole distribution ${\cal P}
[\rho ]$. 

\item{{\em the Satisfiability problem} :} this FC model will be described
in Section V. 
At this stage, we only recall its Hamiltonian based on K-spins interactions~:
\begin{equation}
{\cal H}_{eff} ^{SAT} = - \frac{\alpha}{\beta} \ \ln \left[
\sum _{\vec \sigma _1 , \ldots ,
\vec \sigma _K } c(\vec \sigma _1 ) \ldots
c(\vec \sigma _K ) \  \exp \left( 
- \beta \sum _{a=1} ^n \prod _{\ell =1}^K 
\delta [ \sigma _\ell ^a ; 1]  \right) \right]
\qquad ,
\label{Ksath}
\end{equation}
where $\delta[.;.]$ denotes the Kronecker function \cite{court} and $\alpha$ 
is a positive real parameter. We obtain
\begin{eqnarray}
\frac{1}{n} {\cal H}_{eff} ^{SAT}&=& - \frac{\alpha}{\beta} 
\int _V {\cal D} \rho _1 \ldots {\cal D} \rho _K \; {\cal P}[\rho _1 ]
\ldots  {\cal P}[\rho _K ]  \frac 1m \ln \left[ \int _R dh_1 \ldots dh_K
\right .\nonumber \\ && \left. \rho _1(h_1) 
\ldots \rho _K (h_K) \left( 1 + (e^{-\beta} -1) \frac{ e^{\beta \sum
_{j=1}^ K h_j}}{ \prod _{j=1}^K 2 \cosh \beta h_j } \right)
^m \right] \qquad . 
\label{enerksat}
\end{eqnarray}
Note the additional complexity of (\ref{enerksat}) with respect to
(\ref{enervb}) due to the presence of multispins interactions.
\end{itemize}

\subsection{Entropic contribution}

The calculation of the model-independent entropic contribution
\begin{equation}
S= - \sum _{\vec \sigma } c(\vec \sigma ) \ \ln c(\vec \sigma )
\label{laks}
\end{equation}
to the free-energy is far more complicated. Indeed, contrary to 
${\cal H}_{eff}$, $S$ is not a function of a finite number of the integer
moments
\begin{equation}
C_{\ell} = \sum _{\vec \sigma }  [ c(\vec \sigma ) ] ^\ell 
\end{equation}
of the order parameter. The computation of the entropy $S$ is exposed
in Appendix A. The final expression, in the small $n$ limit 
\begin{eqnarray}
\frac 1n \; S &=& - \int {\cal D} \hat \nu \; {\cal D} \nu \; \exp\left( 
-i \int_{-m}^m dy\; \hat \nu(y) \nu 
(y) \right) c [ i\nu ] \ln c [i \nu ]
\nonumber \\
&& \qquad \frac 1m \; \ln \left[
\int _R \frac{dx}{2\pi}\; \int _{-m}^m dy \;
e^{-i x y} \; (2 \cos  x )^m \; \exp \hat \nu ( y )  \right]
\qquad . \label{entrorsb}
\end{eqnarray}
involves a double functional integrals over $\hat \nu $ and $\nu$ functions,
see Appendix A. The order parameter $P$ enters expression (\ref{entrorsb})
through the $c$ functional (\ref{cnu}) as expected.   

\subsection{Back to RS~: the m=0 and m=1 cases}

In order to be self-consistent, the above formalism has to ensure that
the RS free-energy is found back when $m=0$ or $m=1$. We shall now
see that this is the case.

We first look at the energetic part of ${\cal F}$. 
Within the RS Ansatz, the order parameter $P[\rho ]$ simplifies to 
(\ref{rsfgt}). It is a simple exercise to obtain the resulting expressions
for the energy, see e.g. (\ref{enersk},\ref{enerksat}). 
We observe that the same expressions are found back if we identify
the RS field distribution $p_{rs} (h)$ with

\begin{itemize}
\item{ Case $m=0$ :} 
\begin{equation}
p_0 (h) = \int _V {\cal D}\rho \; P[\rho ] 
\; \rho (h) \qquad . \label{p0}
\end{equation}

More generally, the expansion of the one-step free-energy functional
in powers of $m$ coincides with an expansion in terms
of the cumulants of $P[\rho ]$.  For instance, the first non trivial
term (of order $m$) includes the two functions correlation
\begin{equation}
\Gamma (h,h') = \int _V {\cal D}\rho \; P[\rho ] \; \rho (h) \;\rho
(h')
\end{equation}
which is diagonal in the RS assumption and shows off-diagonal
contributions otherwise \cite{eilat}.

\item{ Case $m=1$ :} 
\begin{equation}
p_1 (h) = \int _V {\cal D}\rho \; P[\rho ] 
\; \delta( h -H[\rho ] )\qquad . \label{p1}
\end{equation}
where
\begin{equation}
H[\rho ] = \frac 1{\beta} \; \tanh ^{-1} \left(
\int _R dh \; \rho (h) \; \tanh \beta h \right) \qquad .
\label{pp1}
\end{equation}
When $m=1$, an exponential number of states with exponentially small 
weights contribute to the partition function. As a result, these
states may be considered as effective micro-configurations and
their cluster as a single effective `state'. 
Equations (\ref{p1},\ref{pp1}) express that the
mean spin magnetisations in the latter are simply the averages of the 
spins magnetisations over all physical states (with vanishingly small weights).

\end{itemize}

The validity of the above relationship between RS theory and the $m=0,1$ 
cases also holds for the entropic part of ${\cal F}$, see Appendix B. 
In addition, we show in Appendix B how to compute the 
path integrals in (\ref{entrorsb}) in the RS scheme to obtain the RS entropy 
\begin{equation}
\frac 1n \; S_{rs} = \int _R \frac{d\hat \nu d\nu}{2\pi} \; e^{-i \hat \nu 
\nu}\; \ln (2 \cosh
\beta \hat \nu )\; c _{rs}(i\nu) \left( 1 - \ln c _{rs} (i\nu) \right) 
\qquad . \label{srs}
\end{equation}
with
\begin{equation}
c _{rs} (i\nu) = \int _R dh\; p_{rs}(h) \; e^{i\nu h}
\qquad .
\end{equation}
The field distribution $p_{rs}$ is obtained through the
optimisation of both entropic and energetic contributions to the
RS free-energy. This approach, that gives the same result as
previous works in the RS framework, we shall now extend to the 
first step of RSB.

\section{Saddle-point equation for the functional order parameter}

We shall now obtain the saddle-point equation fulfilled by the
order parameter $P[\rho ]$. To do so, we differentiate both entropic
and energetic parts of the free-energy functional.

\subsection{Differentation of the entropic part}

The entropy (\ref{entrorsb}) depends on the order parameter through $c$
(\ref{cnu}) only. It results convenient to introduce the 
operator
\begin{eqnarray}
{\cal K} [ \rho , \nu ] &=& 
\int {\cal D} \hat \nu \; \exp \left[ 
-i \int_{-m}^m dy \;\nu(y) \left( 
\hat \nu (y) - \ln \left(
\int _R dh \rho  (h) \frac{ e^{\beta h y }}{(2 \cosh \beta h )^m }
\right) \right) \right]
\nonumber \\
&& \qquad \frac 1m \; \ln \left[
\int _R \frac{dx}{2\pi}\; \int_{-m}^m dy \; e^{-i x y} \; (2 \cos x
)^m \; \exp \hat \nu (y )  \right] \qquad ,
\label{ktens}
\end{eqnarray}
which depends on both functions $\rho (h)$ and $\nu (y )$.
Using the above definition, we rewrite the derivative of the entropic
contribution to the free-energy as
\begin{equation}
\frac 1n \frac{\partial S}{\partial P[\rho ] } =
- \int {\cal D} \nu \; {\cal K} [ \rho , \nu ] \; \ln
c[i \nu ]
\qquad ,
\label{zxc1}
\end{equation}
which depends on the field distribution $\rho$ through (\ref{ktens}).

\subsection{Differentation of the energetic part}

As emphasized in paragraph III.2, the calculation of the (model-dependent)
energetic part of the free-energy, and consequently of its derivative does
not present any difficulty. It turns out that
the latter may always be written under the convenient form 
\begin{equation}
\frac 1n \frac{\partial {\cal H}^{eff}}{\partial P[\rho ] } =
\int {\cal D} \nu \; {\cal K} [ \rho , \nu ] \;
\Omega[i \nu ]
\qquad .
\label{zxc2}
\end{equation}
The expressions of $\Omega$ we now give for the
different models of interest.
\begin{itemize}

\item{{\em the SK model} :} 
\begin{equation}
\Omega ^{SK}[\nu ] = - \frac{\beta}{2} \left( m (1-q_1) \nu _0  +  
q_0 (\nu _1 )^2 + (1-q_1 )\nu _2 \right) \qquad ,
\label{enerskty}
\end{equation}
where $q_0, q_1$ have been defined in (\ref{poiu}) and 
\begin{equation}
\nu _j =  \int _{-m}^m dy \; \nu (y ) \; y ^j
\qquad .
\label{poi1}
\end{equation}

\item{{\em the Viana-Bray model} :} 
\begin{eqnarray}
\Omega ^{VB}[\nu ]&=& -\frac{2 \alpha}{\beta}
\int _V {\cal D} \rho _1 \; {\cal P}[\rho _1 ]
\; \frac 12 \sum _{\sigma = \pm 1}
\exp \left( \int _{-m}^{m} dy\; \nu (y ) 
\right .\nonumber \\ && \left. \ln \left[ \int _R 
dh_1 \;  \rho _1 (h_1 ) \exp \left(
\frac 12 \left( m (A_+ + A_-)  + \sigma \; y
(A_+ - A_-) \right) \right) \right] \right) \nonumber \\
A_{\epsilon} &\equiv & \ln \left[ \frac{ \cosh \beta ( h_1 + \epsilon )}
{\cosh \beta h_1 } \right] \qquad \qquad (\epsilon = \pm 1 ) \qquad . 
\label{enervbder}
\end{eqnarray}

\item{{\em the Satisfiability problem} :} 
\begin{eqnarray}
\Omega ^{SAT}[ \nu ]&=& -\frac{\alpha K}{\beta}
\int _V {\cal D} \rho _1 \ldots {\cal D} \rho _{K-1} \; {\cal P}[\rho _1 ]
\ldots  {\cal P}[\rho _{K-1} ]  \; \frac 12 \sum _{\sigma = \pm 1}
\exp \left( \int _{-m}^m dy\; \nu (y ) 
\right .\nonumber \\ && \left. \ln \left[ \int _R \prod _{j=1}
^{K-1} dh_j \;  \rho _j (h_j ) \exp \left(
\frac 12 \left( m + \sigma \; y
\right) A_{K-1} \right) \right] \right) \nonumber \\
A_{K-1} &\equiv & \ln \left[ 1 + (e^{-\beta} -1) \frac{ e^{\beta \sum
_{j=1}^ {K-1} h_j } }{ \prod _{j=1}^{K-1} 2 \cosh \beta h_j } 
\right] \qquad . 
\label{enerksatder}
\end{eqnarray}

\end{itemize}

\subsection{Self-consistency equation for $P[\rho ]$}

Gathering derivatives (\ref{zxc1}) and (\ref{zxc2}) together,
we obtain the following saddle-point equation
\begin{equation}
\int {\cal D} \nu \; {\cal K} [ \rho , \nu ] \;
\left( - \ln c[ i \nu ] -\beta  \Omega [i \nu ] \right) 
= \lambda _0 \qquad , \label{k56}
\end{equation}
which has to be satisfied for any $\rho (h)$. The Lagrange multiplier 
$\lambda _0$ in (\ref{k56}) is determined through the normalisation of
$P[\rho ]$. It does not depend upon $\rho (h)$ and may be computed
for e.g. $\rho (h) = \delta (h)$. Therefore, the saddle-point
equation (\ref{k56}) simply means that $\Omega [i \nu ] -
\ln c[i \nu ] $ is a zero mode of the operator 
\begin{equation}
{\cal Q} [\rho , \nu ] = {\cal K} [ \rho , \nu ] -
{\cal K} [ \delta (h) , \nu ] \qquad .
\end{equation}
The kernel of ${\cal Q}$ contains all constant functionals $\lambda _1$
due to the normalisation of $\rho (h)$. Therefore, we end up with the
following saddle-point equation
\begin{equation}
\ln c[ \nu ] +\beta \Omega [ \nu ]  
= \lambda _1 \qquad ,
\label{rt1}
\end{equation}
where $\lambda _1$ can be determined for e.g. $\nu (y ) =
0$. From (\ref{cnu}) and (\ref{rt1}), we get $\lambda _1 =
\beta \Omega [ 0 ]$, or equivalently
\begin{equation}
c[ \nu ]
= \exp \left( -\beta \Omega [ \nu ] +\beta  \Omega [ 0 ] 
\right) \qquad .
\label{rt2}
\end{equation}
We now have to find the $\nu$ functions 
for which equation (\ref{rt2}) has to be satisfied. 
$P$ is a distribution of normalised
probabilities $\rho (h)$ merely  by convention. Both $P[\rho (h)]$ and $P[
a(\rho ) \; \rho (h)]$ must lead to the same physics independently
of the irrelevant $a$ factors. This is ensured if 
\begin{equation}
\int _{-m} ^m dy \; \nu (y) 
= 0 \qquad ,
\label{cond1}
\end{equation}
that is if condition (\ref{hdw}) is satisfied.  Consequently, $P[\rho
]$ has to be such that equation (\ref{rt2}) is correct for any
function $\nu (y)$ over the range $-m\le y\le m$ with zero integral
$\nu _0$ (\ref{poi1}). 

\subsection{A simple application~: infinite-connectivity models}

For the SK model, the self consistency equation (\ref{rt2}) is
obtained from (\ref{cnu},\ref{enerskty}) and reads
\begin{eqnarray}
\int _V {\cal D} \rho \ {\cal P} [\rho ] \; && 
\exp \left(  \int _{-m} ^m dy \; \nu (y)  \
\ln \left[ \int _R dh\; \rho (h) \frac{ e^{\beta h y}}
{(2 \cosh \beta h )^m} \right] \right) \nonumber \\
&&= \exp \left[  \frac{\beta ^2}{2} \left( q_0 (\nu _1
)^2 +  (q_1 - q_0)\nu _2 \right) \right]
\end{eqnarray}
for any $\nu$ such that $\nu _0=0$. Solving the above equation, one
finds the SK order parameter
\begin{equation}
{\cal P} ^{SK} [ \rho ] = \int _R d\tilde h \; w_{q_0} (\tilde h
) \ \delta \left[ \rho (h) - \frac{w_{q_1 -q_0} (h -\tilde h) ( 2
\cosh \beta h )^m } {\int _R dh' w_{q_1 -q_0} (h' -\tilde h) ( 2 \cosh
\beta h' )^m } \right]
\label{prsbrs}
\end{equation}
where $w_a (z) = e^{-z^2/2a}/ \sqrt{2 \pi a}$.  In eq. (\ref{prsbrs}),
the symbol $\delta$ denotes the Dirac functional,  {\em i.e.}  the
product over all values of $h$ is omitted for simplicity. As expected,
eqs. (\ref{poiu}) are identical to the usual self-consistent equations
for the RSB overlaps $q_0$ and $q_1$, see eq.~(III.41) in \cite{mpv}. 
Notice that the expression of $\rho$ in (\ref{prsbrs}) is in full 
agreement with the cavity derivation of the effective field
distribution, see eq.~(V.29) in \cite{mpv}. 

\section{Application to the 3-Satisfiability problem}

For generic FC models, the exact resolution of the saddle-point
equation (\ref{rt2}) appears to be extremely difficult. At zero
temperature, the limit indeed of interest for optimization problems,
some analytical simplifications take place.  We shall now see on 3-SAT
how quantitative results may be obtained this way.

\subsection{Presentation of the K-Satisfiability problem}

The satisfiability (SAT) problem is the paradigm of the class of hard
(NP--complete) computational problem arising in complexity theory
\cite{garey}.  A pedagogical introduction to the K-SAT problem, a
version of SAT and some of the current open issues in theoretical
computer science may be found in \cite{hayes}.

K-SAT is defined as follows.  Let us consider $N$ Boolean variables
$\{x_i=0,1\}_{i=1,\ldots,N}$.  Choose randomly $K$ among the
$N$ possible indices $i$ and then, for each of them, a literal that is
the corresponding $x_i$ or its negation $\bar x_i$ with equal
probabilities one half. A clause $C$ is the logical OR of the $K$
previously chosen literals, that is $C$ will be true (or satisfied) if
and only if at least one literal is true.  Next, repeat this
process to obtain $M$ independently chosen clauses
$\{C_\ell\}_{\ell=1,\ldots,M}$ and ask for all of them to be true at
the same time (i.e. we take the logical AND of the $M$ clauses). A
logical assignment of the $\{x_i\}$'s satisfying all clauses, if any,
is called a solution of the K--satisfiability problem.

For large instances ($M,N \to \infty$), K-SAT exhibits a striking
threshold phenomenon as a function of the ratio $\alpha =M/N$ of the
number of clauses per variable.  Numerical simulations indicate that the
probability of finding a solution falls abruptly from one down to zero
when $\alpha$ crosses a critical value $\alpha_c(K)$. Above
$\alpha_c(K)$, all clauses cannot be satisfied any longer.  This
scenario is rigorously established in the $K=2$ case, where $\alpha
_c=1$\cite{exact}.  For $K\ge3$, much less is known; $K(\ge 3)$--SAT
belongs to the class of hard computational problems, roughly meaning
that running times of search algorithms are thought to scale
exponentially in $N$.  Some upper and lower bounds on $\alpha _c(K)$
have been derived\cite{Upper} and numerical simulations have recently
allowed to find precise estimates of $\alpha _c$, e.g. $\alpha _c (3)
\simeq 4.2$\cite{nume}.

In order to study the K--SAT problem, we map it onto a
random diluted systems by introducing some spin variables $S_i=\pm 1\,$ (a
simple shift of the Boolean variables) and a quenched (unbiased) matrix $C
_{\ell ,i}=1$ (respectively $-1$) if $x_i$ (resp. $\bar x_i$) belongs to the
clause $C_\ell$, 0 otherwise. Then the energy--cost function
\begin{equation}
E[C ,S]=\sum_{\ell =1}^M\,\delta \left[ \sum_{i=1}^N\,C _{\ell
,i}\,S_i \; ;\; -K\right] \; \; \;,
\label{energy}
\end{equation}
equals the number
of violated clauses and therefore its ground state (GS) properties
describe the transition from the SAT phase $(E_{GS}=0)$ to the UNSAT
phase $(E_{GS}>0)$. Note that a similar cost function was first
introduced for neural networks by Gardner and Derrida \cite{GD}.

The effective Hamiltonian corresponding to the cost function
(\ref{energy}) was given in (\ref{Ksath}). Previous studies have shown
that the RS theory was able to find back the critical threshold of the
2-SAT problem but became wrong for $K(\ge 3)$--SAT instances
\cite{court,long}. In the
following, we therefore concentrate upon the most interesting
$K=3$ case. We briefly recall the RS solution and then present a first RSB
solution.

\subsection{Saddle-point equation and RS solution}

Following the above-mentioned procedure, the saddle-point equation for
${\cal P}$ reads 
\begin{equation}
\ln c[\nu ] = 3 \alpha 
\int _V {\cal D}\rho _1 {\cal D}\rho _2 \; {\cal P} [\rho _1 ]
{\cal P} [\rho _2 ] \;\left( \frac 12 \; e^{\Phi _-} + 
\frac 12 \; e^{\Phi _+} - 1 \right)
\label{asdd}
\end{equation} 
where (for $\epsilon = \pm 1$)
\begin{eqnarray}
\Phi _\epsilon  = \int _{-m} ^m dy 
\; \nu ( y) && \ \ln  \left\{ \int _R dh_1 dh_2 \;
\rho _1 (h_1 ) \rho _2 (h_2 ) \right. \times  \nonumber \\
&& \times \left. \exp \left[ \frac{m+ \epsilon \;
y}{2}  \ln \left( 1+ \frac{e^{-\beta}
-1 }{(1+e^{-2 \beta h_1} ) (1+e^{-2 \beta h_2} ) }
\right) \right] \right\} \nonumber
\ .
\end{eqnarray}
Eq. (\ref{asdd}) has to be satisfied for all $\nu$ of integral zero.
The ground state properties are obtained by sending $\beta \to \infty$.

Let us briefly recall the RS result \cite{court}. Inserting Ansatz
(\ref{rsfgt}) into (\ref{asdd}), we find that the simplest RS solution
includes half-integer fields only \cite{kanter,court,eilat,rem3}
\begin{equation}
p_{rs} (h) = \sum _{\ell = -\infty } ^\infty e^{-\gamma } I_\ell
(\gamma ) \; \delta \left( h - \frac{\ell}{2} \right)
\label{rsol}
\end{equation}
where $I_\ell$ is the $\ell^{th}$ modified Bessel function and
$\gamma$ is self-consistently determined through 
\begin{equation}
\gamma = \frac {3 \alpha}{4}
\left( 1- e^{-\gamma} I_0( \gamma ) \right)^2 
\label{condrs}
\end{equation}
In addition to the SAT phase solution
$\gamma = E_{GS} =0$, there appears a metastable
solution $\gamma >0, E_{GS} \ne 0$  above $\alpha
= 4.67$, see Fig.~1. This solution becomes thermodynamically stable at
$\alpha =5.18$, well above the `experimental' threshold $\alpha \simeq
4.2$. The latter is indeed thought to coincide with a first-order
spin-glass transition \cite{court,long}. Moreover, the presence of 3-spins
interactions in (\ref{Ksath}) suggests that the one step RSB solution
could be exact for 3-SAT and that $m=1$ at the threshold \cite{pspin} .

\subsection{Simplest RSB solution}

We now turn to the RSB solution. In view of (\ref{rsol}), we first
restrict to distributions $\rho$'s on half-integer fields
$h$. Secondly, the cavity theory teaches us that $\rho$ is biased in
favor of large fields (see Section V.2 in \cite{mpv} for a clear
explanation of this point). For IC models, large $h$'s indeed benefit
from a Boltzmann factor $e^{\beta m |h|}$ as can be seen on
(\ref{prsbrs}).  We thus propose the following form for the
distributions $\rho$ with a non zero weight ${\cal P}[\rho ]$ (for
$m=1$),
\begin{equation}
\rho (h) = \omega (\beta ) \sum _{\ell = -\infty } ^\infty \rho _\ell \; 
\exp \left( \beta \frac{|\ell |}{2}  \right) \; \delta \left( 
h - \frac{\ell}{2} \right) 
\quad ,
\label{scal}
\end{equation}
where $\omega (\beta )$ is a normalization factor and the $\rho _\ell$'s do
not contain exponential terms in $\beta$. Identity (\ref{scal}) is
merely the simplest hypothesis compatible with the saddle-point
equation; we shall come back to this point at the end of the letter.
Sending the temperature to zero, we find that ${\cal P}[\rho ]$ is a
function of two variables only which are computed from $\rho$, namely
the ends of its support $\ell ^{(-)}$ and $\ell
^{(+)}$. This results from the zero temperature expression of the
order parameter
\begin{equation}
\lim _{\beta \to \infty } c[\nu /\beta ] =
\sum _{\ell ^{(-)} \le \ell ^{(+)} } {\cal P}_ {\ell ^{(-)} ,\ell
^{(+)} }  e^{ \nu _- \; \ell ^{(-)} + \nu _+\; \ell ^{(+)}}
\qquad , \label{gs}
\end{equation}
where ${\cal P}_ {\ell ^{(-)} ,\ell ^{(+)} }$ equals the sum of the
weights of the $\rho $ functions having support $[
\ell ^{(-)} , \ell ^{(+)} ]$. Note that (\ref{gs}) depends on $\nu$
through 
\begin{eqnarray}
\nu _- &=& \int _{-1} ^0 dy \; \nu (y)\; y
\nonumber \\
\nu _+ &=& \int _0 ^{1} dy \; \nu (y)\; y
\end{eqnarray}
only, in agreement with the statement that (\ref{scal}) is the simplest non RS
solution to (\ref{asdd}). After some algebra, $\Omega ^{SAT}[\nu ]$ 
(\ref{enerksatder}) can also be shown to depend on $\nu _- ,\nu _+ $ only.
The self-consistent equation for ${\cal P}[\rho ]$ simplifies
into an implicit equation for the matrix ${\cal P}_ {\ell ^{(-)} ,\ell ^{(+)}
}$ which can be easily solved,
\begin{equation}
{\cal P}_ {\ell ^{(-)} ,\ell ^{(+)} } = e^{-\Delta} \; \left(
\frac{\Delta} {2} \right) ^{ \ell ^{(+)} - \ell ^{(-)} } \sum _{ \ell
=\ell ^{(-)}} ^{\ell ^{(+)}} \frac{ e^{-\gamma} I_\ell (\gamma )} {
(\ell ^{(+)} -\ell )! (\ell -\ell ^{(-)})!} \ .
\label{solrsbsat}
\end{equation}
As in the RS solution (\ref{rsol}), $\gamma $ controls the decay of
the probability weights of large fields. The new parameter $\Delta$
sets the width of the supports of the $\rho$'s, that is the magnitude
of the fluctuations of the effective fields from state to state. When
$\Delta =0$, the supports of the $\rho$'s shrink to single points and
the RS solution (\ref{rsol}) is recovered, in agreement with
(\ref{rsfgt}). The two self-consistent equations on $\gamma $ and
$\Delta$ as well as the expression of the corresponding GS energy $E _{GS}$
are given in Appendix C. The results are displayed Fig.~1. In
addition to the trivial solution $\gamma =\Delta =E_{GS}=0$, there appears
another solution for $\alpha > 4.45$ with $0 < \gamma < \Delta $ and a
metastable, {\em i.e.} negative, GS energy. This solution becomes
thermodynamically stable at $\alpha = 4.82$.

\section{Conclusion}

The formalism we have presented in this article has permitted us to
find an explicit RSB solution improving the RS saddle-point. Yet, the
predicted threshold $\alpha _c =4.82$ exceeds the `experimental'
value. This discrepancy may stem from the assumption
(\ref{scal}).  Contrary to the IC case, the payoff for large fields
appearing in $\rho$ might be a non-linear function of $|h|$. If
so, an inspection of the saddle-point equation (\ref{asdd}) shows that
$h$ is not constrained to take half-integer values anymore.  This
refinement scheme is reminiscent of the iterative procedure used in
the RS theory to improve the simplest solution (\ref{rsol}). By
increasing the resolution on the fields, the RS thresholds decreased
from $5.18$ down to $4.60$ \cite{long}. 

However, the meaning of non half-integer fields is far from being
clear. From a physical point of view, the effective fields should be
half-integer valued at zero temperature. Previous studies performed
for the Viana-Bray model have nevertheless shown that the half-integer
field distribution is unstable with respect to longitudinal (within
the RS sector) fluctuations \cite{dom2}.  Then, we face the following
dilemna. The RS saddle-point equation admits a physically sensible but
unstable solution and many other ones, whose significances are
dubious. In this context, the RSB solution presented in the last
Section has a remarkable property. While the supports of the $\rho$'s
with non zero weights contain physical (i.e. half-integer) values
only, the resulting effective distribution $P(h)$ (\ref{p1}) includes
Dirac peaks on the integer multiples of a quarter! More precisely, the
RSB ground state energy is equal to the ground state energy of the RS
solution with fields which are integer multiples of a quarter
\cite{long}. Work is in progress to reach a better understanding of
this puzzling coincidence\cite{fut}.

Acknowledgments~: I am particularly indebted to R.~Zecchina for very
fruitful contributions. I also thank A.~Baldassarri, D.~Dean, S.~Franz 
and I.~Kanter for interesting and motivating discussions. 

\appendix

\section{Calculation of the RSB entropy}

Clearly, the analytical continuation of
the $C_\ell$'s to real $\ell$ in the vicinity of the unity will allow
us to compute the entropy (\ref{laks}),
\begin{equation}
S= -\left. \frac{d }{d\ell } \right| _{\ell =1 } C_{\ell} 
\qquad .
\end{equation}
For integer $\ell$, the moments $C_{\ell}$ are easily expressed from
(\ref{crsb}) as $C_{\ell} \simeq 1 + n \Gamma _\ell $ in the small $n$
limit with
\begin{eqnarray}
\Gamma _\ell &=& 
\int _V {\cal D} \rho _1 \ldots {\cal D} \rho _\ell \; {\cal P}[\rho _1 ]
\ldots  {\cal P}[\rho _\ell ]  \frac 1m \ln \left[ \int _R dh_1 \ldots dh_\ell
\right .\nonumber \\ && \left. \rho _1 (h_1) 
\ldots \rho _\ell (h_\ell) \left( \frac{ 2 \cosh ( \beta \sum
_{j=1}^\ell  h_j )}{ \prod _{j=1}^\ell 2 \cosh \beta h_j } \right)
^m \right] \qquad . \label{tyuijhg}
\end{eqnarray}
In order to achieve an analytical continuation of $\Gamma _\ell$, we 
proceed in two steps. First, consider the argument $A$ of the logarithm in
(\ref{tyuijhg}). Defining
\begin{equation}
\hat \nu ( y ; \rho _1 ,\ldots , \rho _\ell ) = 
\sum _{j=1}^\ell \ln \left[ \int _R dh \rho _j (h) \frac{
e^{ \beta h y }}{(2 \cosh \beta h )^m } \right] \qquad ,
\label{deuxt}
\end{equation}
the latter may be rewritten as
\begin{eqnarray}
A [\hat \nu (y ; \rho _1 ,\ldots , \rho _\ell )] &=& 
\int _R \frac{dx dy}{2\pi}\; e^{-i x y} \; (2 \cosh x
)^m \; \exp \hat \nu (i y ; \rho _1 ,\ldots , \rho _\ell )  
\\ &=& 
\int _R \frac{dx}{2\pi}\; \int _{-m} ^m dy \; e^{-i x y} \; (2 \cos x
)^m \; \exp \hat \nu ( y ; \rho _1 ,\ldots , \rho _\ell )
\qquad ,
\label{deux}
\end{eqnarray}
where the last expression has been obtained through the rotation
$(x,y) \to (ix , -i y)$. Note that the range of integration over $y$ in
(\ref{deux}) makes the integral over $h$ in (\ref{deuxt}) convergent. 
We now introduce the probability distribution ${\cal Q} _\ell [g]$ 
of the functions $g( y ; \rho _1 ,\ldots , \rho _\ell  )$,
\begin{equation}
{\cal Q} _\ell [\hat \nu
] = \int _V {\cal D} \rho _1 \ldots {\cal D} \rho _\ell 
\; {\cal P}[\rho _1 ] \ldots  {\cal P}[\rho _\ell ]  
\ \delta \left[ \hat \nu (y) - \hat \nu ( y ; \rho _1 ,\ldots , \rho _\ell )
  \right]
\qquad , \label{llll}
\end{equation}
and rewrite $\Gamma _\ell$ as a functional integral over all possible 
$g$ functions (with support $[-m,m]$) weighted with measure ${\cal Q} _\ell $,
\begin{equation}
\Gamma _\ell = \int {\cal D} \hat \nu \; {\cal Q} _\ell [\hat \nu ] \
\frac 1m \ \ln A[\hat \nu (y) ] \qquad .
\end{equation}
The second step of the calculation lies in the analytical continuation of
the measure ${\cal Q} _\ell $ to real $\ell$. This may be achieved through
an exponential representation of the functional Dirac in (\ref{llll}). We 
thus obtain
\begin{equation}
\left. \frac{d}{d\ell } \right| _{\ell =1 } {\cal Q} _\ell [\hat \nu] 
= \int {\cal D} \nu \; \exp\left(- i \int_{-m}^m dy \hat \nu (y) 
\nu (y) \right) c [ \nu ] \ln c [ \nu ]
\label{dqldl}
\end{equation}
where the functional $c[ \nu ]$ is given in (\ref{cnu}).
In equation (\ref{dqldl}), the measure ${\cal D} \hat \nu$ ensures the correct 
normalisation of the Dirac functional, that is factors $1/2\pi$ have to be 
included when discretising the path integral over $\hat \nu$. The resulting
expression of the entropy is given in (\ref{entrorsb}).

\section{RS entropy and the $m=0,1$ cases}

In this Appendix, we first derive the RS expression of the entropy 
(\ref{entrorsb}). We then show how the latter is recovered when $m=0$ or 
$m=1$.

\subsection{Calculation of the RS entropy}

Consider the RS expression (\ref{rsfgt}) for $P[\rho ]$. 
The functional $c[ \nu ]$ (\ref{cnu}) simplifies to 
\begin{equation}
c( \nu_0 , \nu_1 ) = \int _R dh\; p_{rs} (h) \exp
\left( -i\; \nu_0\; m \;\ln ( 2 \cosh \beta h ) +i\; 
\nu _1\; \beta \; h \right)
\qquad .
\end{equation}
It depends on $\nu$ through the
two moments $\nu _0 , \nu _1 $ only, see (\ref{poi1}).
We now introduce the series expansion of $\hat \nu (y)$ around $y =0$,
\begin{equation}
\hat \nu (y) = \sum _{j=0} ^\infty \hat \nu _j \; y ^j
\qquad .
\label{poi2}
\end{equation}
Combining the coefficients definitions (\ref{poi1}) and (\ref{poi2}), we
rewrite the coupling term between $\hat \nu $ and $\nu$ in (\ref{entrorsb})
as follows
\begin{equation}
\exp \left(- i\int _{-m}^m dy \; \hat \nu (y )\nu (y ) \right) =
\exp \left( -i
\sum _{ j =0}  ^\infty \hat \nu _j \; \nu _j  \right)\qquad .
\end{equation}
The integration over all coefficients $\nu_j$ with $j\ge 2$ gives
$\hat \nu _j =0$, $\forall j\ge 2$. Therefore, only the $\hat \nu $'s 
that are linear functions of their argument $y$ survive. We obtain
\begin{eqnarray}
\frac 1n S &=& - \int _R \frac{d\hat \nu _0 d\nu_0}{2\pi} \;
\frac{d\hat \nu _1 d\nu_1}{2\pi} \; c( \nu_0 , \nu_1 ) \ln
c( \nu_0 , \nu_1 ) e^{-i ( \hat \nu _0 \nu_0 + \hat \nu _1 \nu_1 )}
\nonumber \\
&&  \frac 1m \; \ln \left[ \int _R \frac{dx dy}{2\pi} \;
e^{-i x y} \; (2\cosh x )^m \; \exp ( \hat \nu _0 + 
\hat \nu _1 \; i y )
\right] \qquad .
\label{asd1}
\end{eqnarray}
The argument of the logarithm in (\ref{asd1}) simply reads
$e^{\hat \nu _0}\; ( 2\cosh  \hat \nu _1 )^m$. 
The remaining integrals on $\hat \nu _0, \nu _0, 
\hat \nu _1 , \nu _1$ may then
be carried out and the final result is given in (\ref{srs}).

\subsection{Case m=0}

We expand the logarithm in (\ref{entrorsb}) and find
\begin{eqnarray}
\frac 1m \; \ln \left[
\int _R \frac{dx dy}{2\pi}\; e^{-i x y} \; (2 \cosh x
)^m \; \exp \hat \nu  (i y )  \right] &=& \nonumber \\
\frac{\hat \nu (0)}{m} + \int _R  \frac{dy d\hat y}{2\pi}\; 
e^{-i y \hat y} \; \ln
(2 \cosh y ) \; e^{ \hat \nu  (i \hat y ) - \hat \nu (0)} + O(m) &&
\qquad , \label{expanlog}
\end{eqnarray}
in the small $m$ limit. We now consider the first term on the
r.h.s of (\ref{expanlog}) and integrate out all modes $\hat \nu (y \ne
0)$ in (\ref{entrorsb}). The resulting $c [\nu ]$ reads, for small $m$,
\begin{equation}
c [\nu ] = 1 + i\; m\; \nu (0) \int _R dh\; p_0 (h) \; \ln (2
\cosh \beta h ) + O(m^2) \qquad ,
\end{equation}
where $p_0$ has been defined in (\ref{p0}). Integrating
$g(0),\nu(0)$ out, we obtain a first contribution to the entropy 
(\ref{entrorsb}),
\begin{equation}
\frac 1n \; S_1 = \int _R dh \; p_0 (h) \; \ln (2
\cosh \beta h )
\qquad .
\label{ss1}
\end{equation}
We now focus on the second part of the r.h.s of (\ref{expanlog}). 
By integrating the $\hat \nu $'s out, we get $\nu (y )=0$ except
$\nu (i \hat y) = - \nu (0) = i$. The corresponding functional
(\ref{cnu}) reads 
\begin{equation}
c [\nu ] \equiv c (\hat y )
= \int _R dh \; p_0 (h) \; e ^{i \beta h \hat y } \qquad ,
\end{equation}
giving rise to the following entropy
\begin{equation}
\frac 1n \; S_2 = - \int _R  \frac{dy d\hat y}{2\pi}\; e^{-i y \hat y}
\; \ln (2 \cosh y ) \; c (\hat y) \; \ln c (\hat y )
\qquad .
\label{ss2}
\end{equation}
Summing up $S_1$ and $S_2$, the RS entropy (\ref{srs}) is recovered with
$p_{rs}=p_0$.

\subsection{Case m=1}

For $m=1$, the argument of the logarithm in (\ref{entrorsb}) reads
$e^{\hat \nu 
_+} +e^{\hat \nu _-}$, with $\hat \nu 
_\epsilon = \hat \nu (i \epsilon )$, $\epsilon = \pm 1$. 
For all $y \ne \pm 1$, the integration
over $\hat \nu (y)$ gives $\nu (y) =0$. The entropy then reads
\begin{equation}
\frac 1n S = - \int _R \frac{d\hat \nu _+ d\nu_+}{2\pi} \;
\frac{d\hat \nu _- d\nu_-}{2\pi} \; c( \nu_+ , \nu_- ) \ln
c( \nu_+ , \nu_- ) e^{i ( \hat \nu _+ \nu _+ + \hat \nu _- \nu _- )}
\; \ln \left[ e^{\hat \nu 
_+} +e^{\hat \nu _-} \right] \qquad ,
\label{asd2}
\end{equation}
where, using definition (\ref{p1}),
\begin{equation}
c( \nu_+ ,\nu_- ) = \int _R dh\; p_1 (h) \exp
\left( i\; (\nu_++\nu_-) \;\ln ( 2 \cosh \beta h ) - i\; 
(\nu_+-\nu_-)\; \beta h \right)
\qquad .
\end{equation}
We make the following unitary change of variables
\begin{eqnarray}
\hat \nu _0 &=& \frac 12 (\hat \nu _+ + \hat \nu _-) \nonumber \\
\nu_0 &=& \nu_+ + \nu_- \nonumber \\
\hat \nu _1 &=& \frac 12 (\hat \nu _+ - \hat \nu _-) \nonumber \\
\nu_1 &=&  \nu_+ - \nu_- 
\qquad ,
\end{eqnarray}
and find back the RS expression (\ref{asd1}). Therefore, when $m=1$,
the RS entropy is recovered with $p_{rs}=p_1$.

\section{Self-consistency equations for $\gamma, \Delta$ and
Ground-state energy $E$}

The self-consistent equations for $\gamma$ and $\Delta$ read 
\begin{eqnarray}
\gamma &=& 3\alpha \left( \frac 12 -\frac{r_0}{2} -r_1
\right)^2 \label{fin1} \\ 
\Delta &=& 3\alpha \left( \frac 12 -\frac{r_0}{2} \right)^2
- \gamma \qquad ,
\label{fin2}  
\end{eqnarray}
where 
\begin{eqnarray}
r_0 &=& \sum _{\ell = -\infty } ^\infty e^{-\gamma - \Delta }\;
I_{\ell} (\gamma )\; I_{2\ell} (\Delta ) 
\nonumber \\ 
r_1 &=& \frac 12 \sum _{\ell = -\infty } ^\infty e^{-\gamma - \Delta }\;
I_{\ell} (\gamma )\; \left[ I_{2\ell -1} (\Delta ) +
I_{2\ell +1} (\Delta ) \right]
\qquad .  
\end{eqnarray}

To compute the ground state energy $E$, we insert the saddle-point
solution for $c(\vec \sigma )$ into the free-energy (\ref{e2}) and
keep the linear term in $n$ only. We then send $\beta \to \infty$ and
find
\begin{eqnarray}
E &=& -\alpha \left[ \left( \frac{1-r_0}{2} \right) ^3 +
\left( \frac{1-r_0}{2} -r_1 \right) ^3 \right] + 
\frac{3 \alpha}{4} \left[ \left( \frac{1-r_0}{2} \right) ^2 +
\left( \frac{1-r_0}{2} -r_1 \right) ^2 \right] - \nonumber \\
&&\frac{\Delta }{4} (r_0 +r_1 ) - \frac{\gamma}{4} 
(r_0 +r_2 +2 r_1 ) \nonumber \\
r_2 &=& \frac 12 \sum _{\ell = -\infty } ^\infty e^{-\gamma - \Delta }\;
I_{\ell} (\gamma )\; \left[ I_{2\ell -2} (\Delta ) +
I_{2\ell +2} (\Delta ) \right]
\label{tyur}
\qquad .  
\end{eqnarray}
Note that $\Delta =0 $ is always a solution of the self-consistent
equation (\ref{fin2});  eq. (\ref{fin1}) then
gives back the RS saddle-point constraint on $\gamma$
(\ref{condrs}). In this case, the ground state energy $E$ (\ref{tyur})
simplifies to
\begin{equation}
E_{RS} = \frac{\gamma}{6} \left( 1- e^{-\gamma} I_0(\gamma )- 
3 e^{-\gamma} I_1(\gamma ) \right)
\qquad ,
\end{equation}
in agreement with the findings of \cite{court,long}.

\begin{figure}
\epsfxsize=468pt\epsffile{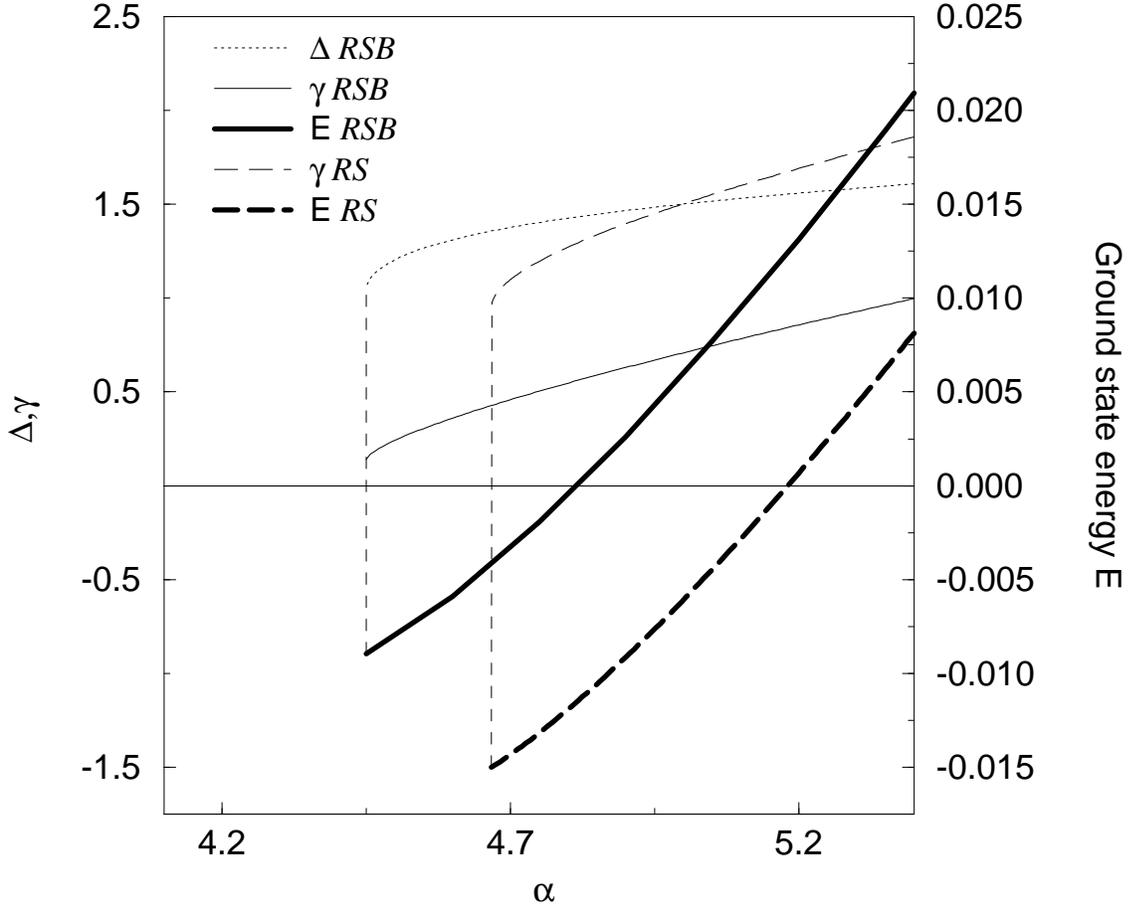}
\vskip 1cm
\caption{Decay parameters $\gamma$ (thin curves-left side scale) and
ground state energies $E$ (bold curves-right side scale) for the RSB
(full curves) and the RS (long-dashed curves) solutions as functions
of the number of clauses per variables $\alpha$. The RSB width
parameter $\Delta$ is shown on the upper dotted curve. The vertical
dashed lines indicate the spinodal points $\alpha_{RSB}=4.45$ and
$\alpha _{RS} = 4.67$.}  
\protect\label{f1}
\end{figure}


\begin{thebibliography}{99}
\bibitem{rem1}
Another generating function of all overlaps was already introduced by
Mottishaw and De Dominicis for technical reasons.  Their order
parameter however includes some disorder dependence (see equation (3)
in \cite{dom1}) and its interpretation is therefore less clear than
$c(\vec \sigma )$'s one. See also \cite{gold}.
\bibitem{clown}
For IC models, the only non trivial moment appearing in $c[\nu ]$ is
the variance $\int _{-m} ^ m dy\; \nu (y)\; y^2$. This oversimplification
reflects the Gaussian nature of the distribution in (\ref{prsbrs}).
\bibitem{rem3}
The field $h$ equals half the difference between the numbers of 
violated clauses when flipping a Boolean variable (see Section IX
in \cite{court}). It may thus take half-integer values \cite{eilat}.
\end{thebibliography}
\end{document}